\documentclass{article}

 \usepackage{graphicx} \usepackage{amsmath}

\suppressfloats[t]

\begin{document} \large

\begin{center}
{\bf Jet formation in the interaction of localized waves on the free surface of dielectric liquid in a tangential electric field}\\[1.0ex]

\vspace{2mm}
{\it Evgeny A. Kochurin$^1$, Nikolay M. Zubarev$^{1,2}$ }\\[1.0ex]
{$^1$Institute of Electrophysics, Ural Division,
Russian Academy of Sciences, 106 Amundsen Street, 620016 Ekaterinburg, Russia\\
$^2$P. N. Lebedev Physical Institute, Russian Academy of Sciences,
53 Leninskij prospekt, 119991 Moscow, Russia}
\end{center}

\vspace{2mm}
\begin{abstract}

\large
Nonlinear dynamics of the free surface of finite depth non-conducting fluid with high dielectric constant subjected to a strong horizontal electric field is considered. Using the conformal transformation of the region occupied by the fluid into a strip, the process of counter-propagating waves interaction is numerically simulated. The nonlinear solitary waves on the surface can separately propagate along or against the direction of electric field without distortion. At the same time, the shape of the oppositely traveling waves can be distorted as the result of their interaction. In the problem under study, the nonlinearity leads to increasing the waves amplitudes and the duration of their interaction. This effect is inversely proportional to the fluid depth. In the shallow water limit, the tendency to the formation of a vertical liquid jet is observed.

\end{abstract}

\vspace{2mm}

It is well known \cite{Me3} that an external electric field directed tangentially to the unperturbed free surfaces, or interfaces of dielectric liquids has a stabilizing effect on the boundary. On the other hand, the normal field results in aperiodic growth of the boundary perturbations (the so-called electro-hydrodynamic instability) \cite{Me4}. In recent years, new experimental results concerning the influence of electric field on dynamics of capillary waves on the surface of dielectric liquids have been obtained \cite{exper1,exper2}. At the present time, the nonlinear structures formation on the interface of fluids in a vertical electric field is well studied (see, e.g., \cite{kuz,zuf,ko1,ko2}). The interest in studying the nonlinear dynamics of fluids interfaces under the action of an electric field is caused by the possibility of controlling and suppressing of boundary instabilities. The stabilization of Rayleigh--Taylor and Kelvin--Helmholtz instabilities (for technological processes they can be considered as undesirable effects \cite{laser1, laser2}) by the external electric field was investigated in \cite{Kor,RT} and \cite{El,Elh,zu0}, respectively. In general, the stabilization of hydrodynamic instabilities is a complex problem for which solution it is necessary to understand how the liquid boundary behaves in a strong horizontal electric field without destabilizing factors.

In the present work, the nonlinear dynamics of the free surface of finite depth layer of a non-conducting liquid with high dielectric constant is studied in the situation where the tangential electric field is the only factor determining the fluid motion. In the case of high fluid permittivity, the normal (destabilizing) component of the field is much smaller than the tangential (stabilizing) one. In such a situation, nonlinear localized waves on surface can propagate along or against the electric field direction without distortion, i.e., the equations of motion admit a wide class of exact traveling wave solutions \cite{zu1,zu2,zu4}. The methods of computer simulation based on the dynamic conformal transforms of the region occupied by the fluid into the canonical region (a strip) were applied. At present time, the computational techniques based on the conformal transformations develop intensively; see, for example, \cite{conf1, conf2,conf3}.
The simulations show that the counter-propagating spatially localized waves can deform in result of their collisions. The deformation is inversely proportional to the fluid depth. As a consequence, in the case of small depth, a spatially narrow perturbation of the surface appears (its amplitude becomes comparable with the layer depth). This phenomenon should be taken into account in developing the methods of controlling and stabilizing the hydrodynamic instabilities.

\section{Initial equations}
Let us consider the nonlinear dynamics of the free surface of dielectric liquid with finite depth $d$ in an external horizontal electric field. In the unperturbed state, the boundary of the liquid is the horizontal plane $y=0$ (the $x$ axis of the Cartesian coordinate system lies in this plane and the $y$ axis is perpendicular to it). Let the electric field be directed along the $x$ axis and be $E$ in magnitude. The problem under consideration is periodic along $x$ axis, i.e., $E=U/l$, where $U$ is the voltage drop at the space period $l$. Let the function $\eta (x,t)$ specify the deviation of the boundary from the plane, i.e., the equation $y=\eta $ determines the profile of the surface. Due to the mass conservation law, the mean depth is constant,
$$\frac{d}{d t}\int\limits_{-l/2}^{l/2} \eta (x,t)d x=0.$$

We assume that fluid is inviscid and incompressible, and its flow is irrotational (potential). The velocity potential $\Phi$ satisfies the Laplace equation,
\begin{equation}\label{lap}\nabla^2\Phi =0,\qquad-d< y<\eta,\end{equation}
with the condition $\Phi_y=0,$ at bottom $y=-d$.
Here and below $\nabla=\{\partial_x,\partial_y\}$ is the two-dimensional nabla operator. The kinematic and dynamic boundary conditions at the free surface have the form,
\begin{equation} \label{kin} \eta_t=\Phi_y-\eta_x\Phi_x, \qquad y=\eta,\end{equation}
\begin{equation} \label{dyn}\Phi_t+\frac{1}{2}\left(\nabla\Phi \right)^2=\varepsilon_0 \varepsilon\frac{ (\nabla\varphi)^2-E^2}{2\rho},\qquad y=\eta,\end{equation}
where $\varphi$ is the electric field potential, $\rho$ is the fluid density, $\varepsilon_0$ is the vacuum permittivity, $\varepsilon\gg1$ is the dielectric constant of the liquid. The equations (\ref{lap})-(\ref{dyn}) should be complemented by a system of equations for the electric field potential.

In the absence of free electric charges, the electric field potential satisfies the Laplace equation
\begin{equation*}\nabla^2\varphi =0,\qquad-d< y<\eta.\end{equation*}
We consider the liquid with high dielectric constant. Then, according to \cite{field}, it is possible to solve the problem of the field distribution in the fluid regardless the field distribution outside of it ($y>\eta$, $y<-d$). The electric field potential obeys the conditions at free surface and, respectively, at the bottom,
\begin{equation*}\varphi_y-\eta_x\varphi_x=0,\quad y=\eta;\qquad \varphi_y=0, \quad y=-d.\end{equation*}
These expressions imply that the field is directed tangentially to both boundaries.

The system under study is conservative. Its total energy,
\begin{equation*}H=\frac{\rho}{2}\int {\left(\nabla\Phi\right) ^2d^2r}-\frac{\varepsilon_0\varepsilon}{2}\int {\left[\left(\nabla\varphi\right) ^2-E^2\right]d^2r},\end{equation*}
does not change with time.

The relations presented above are complete equations system describing the motion of the finite depth layer of a non-conducting fluid with high dielectric constant under the action of electrostatic forces caused by the external horizontal electric field. Influence of capillary and gravitational forces is not taken into account, which corresponds to the limit of a strong field. In the linear approximation, the system of governing equations describes the plane waves propagation with the following dispersion relation: $\omega^2=c^2k^2$, where $\omega$ is the frequency, $k$ is the wavenumber, $c=(\varepsilon_0\varepsilon E^2/\rho)^{1/2}$ is the velocity of linear waves.

For convenient further consideration, we pass to the dimensionless variables as
\begin{equation*}
\Phi \to \lambda c\,\Phi,
\qquad
\varphi \to \lambda E\,\varphi,
\qquad
x \to \lambda\,x,
\qquad
\eta \to \lambda\,\eta,
\qquad
t \to \lambda c^{-1}\,t,
\end{equation*}
where $\lambda$ is the characteristic wavelength. In the new variables, the velocity of linear waves is unity.

\section{Conformal variables}
By analogy with \cite{zu4,fin,ruban,dut}, we make the conformal transformation of the region occupied by the liquid into the parametric strip
\begin{equation*}-D<v\leq0, \qquad-\infty<u<+\infty,\end{equation*}
where $D=D(t)$ is the conformal depth of fluid, which, in the general case, can depend on time. The fluid free boundary and the bottom
correspond to $v=0$ and $v=-D$, respectively. The auxiliary variables $u$ and $v$ in the problem under study have clear physical meaning: $u$ coincides with the field potential $\varphi$ except for the sign and the condition $v = \mbox{const}$ specifies the electric field lines. In the new variables, the Laplace equations for the electric field potential and velocity potential can be solved analytically. As a result, the initial (2+1)-dimensional problem of the liquid motion can be reduced to the problem of motion of its free surface, which has a lower dimension of (1+1). In the conformal variables, the problem is also periodic with respect to $u$ with the same period $l$.

The surface of the liquid in the new variables is specified by the parametric expressions
\begin{equation*}y=Y(u,v=0,t),\qquad x=X(u,v=0,t)=x_0(t)+u-\hat T Y, \end{equation*}
where $x_0(t)$ is a function of time determining the origin of the $X$-coordinate in the physical domain, $\hat T$ is an integral operator; the action of $\hat T$ and of its inverse operator $\hat R$ is defined in Fourier space as
\begin{equation*}\hat T_k=i \coth(kD),\qquad \hat R_k=-i \tanh(kD).\end{equation*}
The conformal depth $D$ is related with the physical depth $d$ as follows \cite{fin}
\begin{equation*} D(t)=d+\frac{1}{l}\int\limits_{-l/2}^{l/2}Y(u,t)du.\end{equation*}
It should be noted that for small amplitude waves the deviation of conformal depth from the physical one is small, i.e., $D\approx d$.

In the present work, we will use the results of \cite{zu4, fin,ruban,dut} to obtain the equations in conformal variables. In the new variables, the kinematic boundary condition (\ref{kin}) takes the form
\begin{equation} \label{eq1}
Y_t=q(t)Y_u+\left[1-\hat T Y_u\right]\frac{\hat R \Psi_u}{J}-Y_u \hat T\left(\frac{\hat R \Psi_u}{J}\right),
\end{equation}
where $q(t)$ is a function of time related to $x_0(t)$, for more details, see \cite{fin}, $\Psi$ is the function defining the velocity potential at the surface, and $J=X_u^2+Y_u^2$ is the Jacobian of conformal transform (its inverse value corresponds to the electrostatic pressure at the boundary $P_E=1/2J$). The dynamic boundary condition is rewritten as
\begin{equation} \label{eq2}
\Psi_t=q(t)\Psi_u-\frac{1}{2J}\left[\Psi_u^2-\hat R \Psi_u^2\right]-\Psi_u \hat T\left(\frac{\hat R \Psi_u}{J}\right)+\frac{1-J}{2J}.
\end{equation}
The last term in right-hand side of this equation is responsible for the action of the tangential electric field. Due to the Galilean invariance of the system (\ref{eq1})-(\ref{eq2}), one can put $q(t)\equiv0$ without loss of generality.

The pair of equations (\ref{eq1}) and (\ref{eq2}) completely describes the dynamics of the free surface of an ideal non-conductive fluid with finite depth in a strong tangential electric field. In the limit $d \to +\infty$, the system (\ref{eq1})-(\ref{eq2}) turns into the equations system describing a free surface of deep dielectric liquid \cite{ko}.

\section{Simulation of waves interaction}
The localized waves play an important role in the study of propagation and interaction of nonlinear surface waves. The collision between oppositely propagating solitary waves on the surface of deep liquid was simulated in \cite{ko}. It was shown that the interaction of localized waves is elastic, i.e., the energy and the horizontal momentum of each of the oppositely propagating interacting solitary waves are conserved. Nonlinearity results in deformation of the surface waves, as consequence, regions with steep wave front are formed. In the same time, the solitary waves remained structurally stable, i.e., they did not radiate additional wave packets. Thus, the question arises whether the nonlinear waves on the surface of finite depth fluid have the same property and what is the influence of bottom on the system dynamics.

An effective way to answer this question is to integrate the system (\ref{eq1})-(\ref{eq2}) numerically. In order to solve the equations (\ref{eq1}) and (\ref{eq2}), it is convenient to use the pseudo-spectral methods of calculations, i.e., to approximate all functions by the finite Fourier series. In Fourier space, the differentiation procedure has the form $\partial_u f \to ik f_k$, where $f_k$ is the Fourier image of function $f$. The main difficulty of numerical integration of the equations system (\ref{eq1})-(\ref{eq2}) is calculation of the integral operators $\hat R$ and $\hat T$, because the value of the parameter $D$ can depend on time. The function $D=D(t)$ is determined from the law of mass conservation
\begin{equation}\label{mass}\frac{d M}{d t}=0,\qquad M=\frac{1}{l}\int\limits_{-l/2}^{l/2}\left(1-\hat T Y_u\right)Ydu.\end{equation}
The relation (\ref{mass}) is a nonlinear integral equation with respect to the parameter $D(t)$, which can be resolved by iterations. The numerical integration of the system (\ref{eq1})-(\ref{eq2}) with respect to time will be carried out by the explicit Runge--Kutta method of the fourth-order accuracy.

Let us consider the interaction between two counter-propagating nonlinear localized waves of the following form:
\begin{equation*}Y=Y^++Y^-,\qquad Y^{\pm}=a^{\pm}\exp(-2[u\pm u_0]^2),\end{equation*}
where the symbols $\pm$ correspond to positive and negative direction of the wave propagation along the $x$ axis, $a^{\pm}$ are the waves amplitudes, $\pm u_0$ are the initial positions of the amplitudes maxima. The initial condition for the velocity potential should be taken as follows
\begin{equation*}\Psi=\Psi^{+}+\Psi^{-},\qquad \Psi^{\pm}=\pm \hat T Y^{\pm},\end{equation*}
where the functions $\Psi^{\pm}$ define velocities of the oppositely traveling waves. For the numerical experiments presented below the space period is chosen to be $4u_0$, the time interval is $2u_0$ with the step of discretization $10^{-4}$, and the number of used Fourier harmonics is equal to $8192$.

\begin{figure}[t]
\center{\includegraphics[width=1.1\linewidth]{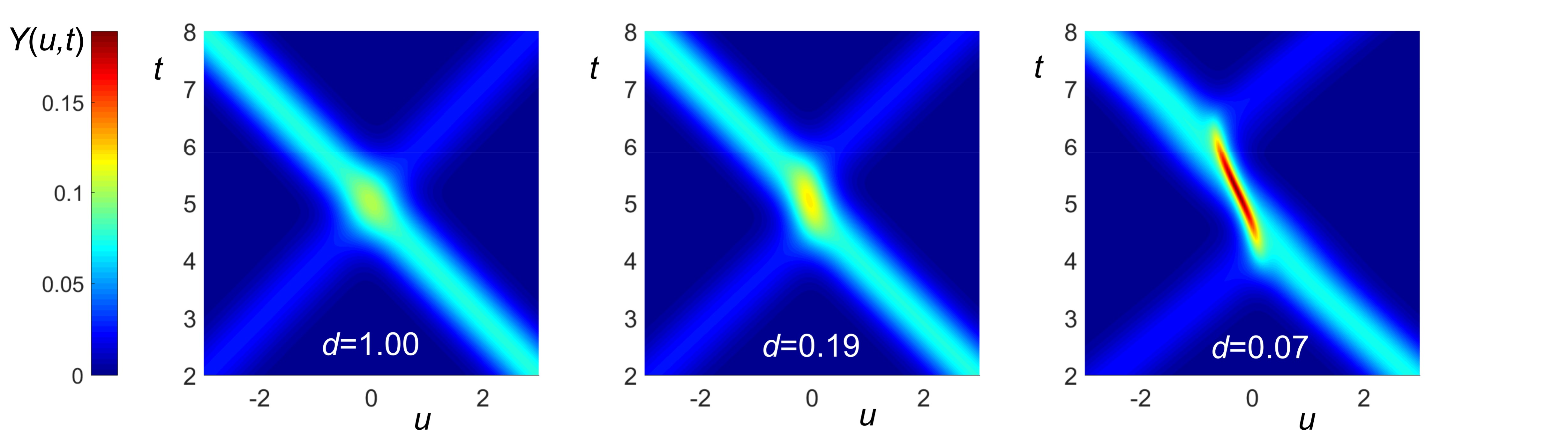}}
\caption{The localized waves interaction for the different values of fluid depth. The calculation parameters are $a^+=0.025,\, a^-=0.075,\, u_0=5,\, l=20$. }
\label{fig1}
\end{figure}

The figure~\ref{fig1} demonstrates the interaction of weakly nonlinear waves for three different values of the fluid depth. For relatively large value of the depth $D_0=D(t=0)=1.006$, which corresponds to $d\approx1$, the interaction is almost linear. Further, the maximum of surface elevation at the collision moment increases as the depth decreases ($D_0=0.2,\,d\approx0.19$), and in the limit of the small depth ($D_0=0.08,\,d\approx0.07$) it exceeds the sum of the amplitudes of waves almost in two times. The increase in surface elevation $\Delta a$ (at the moment of waves collision) versus the fluid depth $d$ is shown in figure~2~(a). The curve {``}1'' corresponds to the numerical simulation results and the curve {``}2'' corresponds to approximate analytical solution for weakly nonlinear waves \cite{zu4}
\begin{equation}\label{sol}Y=Y^++Y^--Y^{+}_u\hat T Y^{-}-Y^{-}_u\hat T Y^+-\hat R(\hat T Y^+ \hat T Y^-)_u.\end{equation}
The functions $Y^+$ and $Y^-$ have a meaning of exact solutions of the linear wave equation; their nonlinear superposition is the approximate solution of the equations system (\ref{eq1})-(\ref{eq2}). It can be seen from the figure~2~(a) that the analytical solution demonstrates a good agreement with simulation results for high values of fluid depth. The amplitude $\Delta a$ increases as the depth decreases growing infinitely at $d \to 0$. In fact, we observe a tendency to the formation of a spatially narrow perturbation on the fluid surface (a vertical jet).

The figure~\ref{fig1} also shows that nonlinearity leads to increase in the interaction duration, i.e., a phase shift arises in result of the waves collision. The effect is stronger for the wave with smaller amplitude. The dependence of the moment of waves collision $\tau$ on depth is shown in the figure~2~(b). Here, we should note that the event of waves collision is defined as the moment of time, for which the potential energy of the system has a maximum and, hence, the kinetic energy has a minimum (for the linear interaction $\tau_l=5$). It is important that the phenomenon of the phase shift cannot be described in the framework of the analytical solution (\ref{sol}).
This is probably related with the fact that the solution (\ref{sol}) was obtained for the spatially localized functions; in the present study the periodicity can affect sufficiently on the system evolution. As it is seen from the figure~2~(b), the effect is inversely proportional to the fluid depth $\tau-\tau_l\sim d^{-1}$, i.e., it is very sensitive to variation of $d $. It is interesting to note that the similar phenomenon of phase shift is demonstrated by the soliton solutions of the Korteweg--de Vries equation \cite{solit}. It is well known that the Korteweg--de Vries solitons remain their initial shape after collision.
Let us also consider the waves shape after interaction in the problem under study.

\begin{figure}[t]
\center{\includegraphics[width=0.8\linewidth]{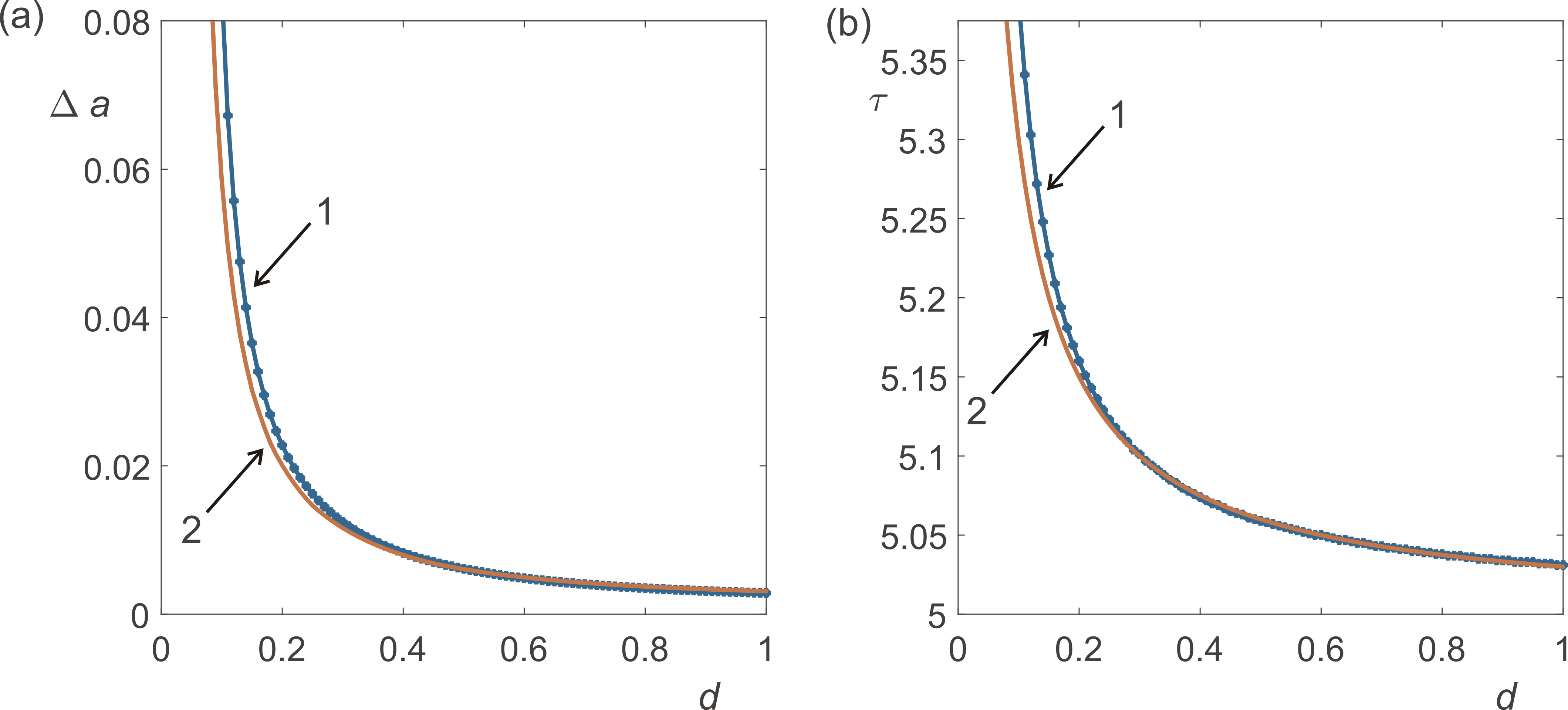}}
\caption{
(a) The amplitude increase $\Delta a$ at the collision moment versus fluid depth; the simulation results (blue curve {``}1'') and approximate analytical solution (red curve {``}2''). (b) The collision moment versus fluid depth; the simulation results (blue curve {``}1''), the power law~$\tau-\tau_l \sim d^{-1}$ (red curve {``}2''). The calculation parameters are $a^+=0.025,\, a^-=0.075,\, u_0=5,\, l=20$.}
\label{fig2}
\end{figure}

The figure~3~(a) shows the spectrum of $Y(u)$ at the initial moment and at the end of calculation interval, i.e., after a single collision of waves. It can be seen that after the interaction, there are additional small-scale harmonics in the spectrum, whose amplitudes exceed the computing error. The spectrum {``}beatings'' near $10^{-16}$ are caused only by computing with double precision. Since the total energy of the system is constant, it is possible to estimate the accuracy of computational methods used. Figure~3~(b) shows relative variations in energy and mass of the system. As can be seen, the calculation error during the simulation period does not exceed the value of $10^{-9}$, i.e., it is extremely low. Thus, in contrast to the interaction of classical solitions, the shape of investigated localized waves changes after interaction.

\begin{figure}[t]
\center{\includegraphics[width=0.8\linewidth]{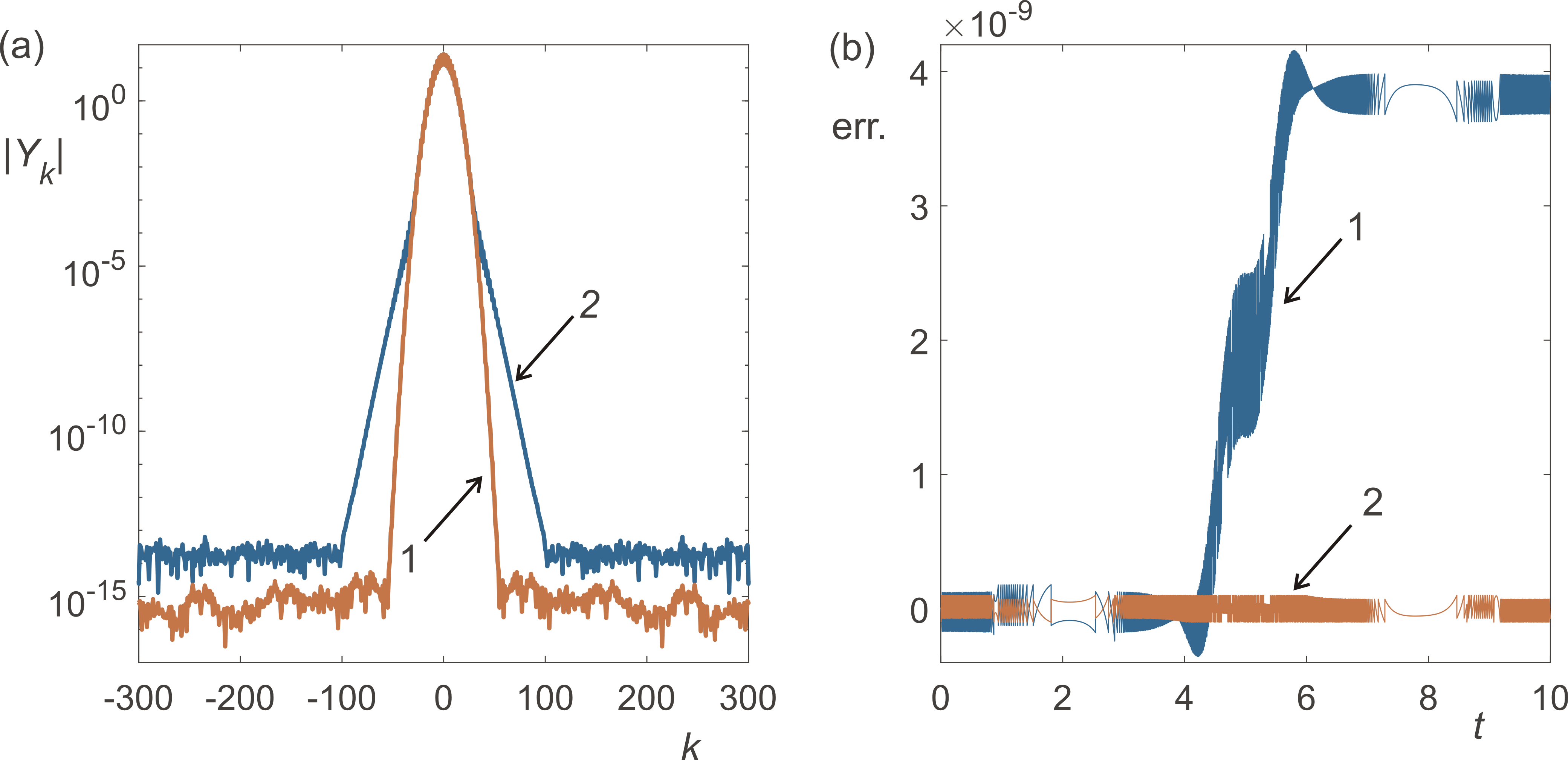}}
\caption{
(a) The spectra of $Y(u,t)$ at the initial moment (blue curve {``}1'') and at the end of calculation interval (red curve {``}2''). (b) The relative variations of energy (blue curve {``}1'') and mass of the system (red curve {``}2''). The calculation parameters are $d=1,\, a^+=0.025,\, a^-=0.075,\, u_0=5,\, l=20$.}
\label{fig3}
\end{figure}

\section{Conclusion}
In the present work, the nonlinear dynamics of the free surface of an ideal dielectric liquid with finite depth in a strong horizontal electric field was numerically simulated. The computational algorithm is based on the conformal transform of the region occupied by the fluid into a strip. The spatially localized waves can propagate separately along, or against the electric field direction without distortion. Our simulations show that their shape can deform in the result of collisions of the counter-propagating waves. The intensity and duration of the nonlinear waves interaction increases as the fluid depth decreases. In the limit of small depth, this leads to the formation of a spatially narrow perturbation of the surface, whose amplitude is comparable or even greater than the depth of the liquid. In other words, there is a tendency to the formation of liquid jets at the free boundary as a result of interaction between oppositely propagating nonlinear waves.
The effect of surface tension on the jet can lead to the ejection of electrically polarized droplets.

\section*{Acknowledgments}
This work was supported by state contract No.\,0389-2014-0006, Russian Foundation for Basic Research (projects No.\,16-38-60002, 16-08-00228, and 17-08-00430), by the Presidential Programs of Grants in Science (No.\,SP-132.2016.1), and by the Presidium of the Ural Branch of the Russian Academy of Sciences (project No.\,15-8-2-8).

\end{document}